\documentclass[fleqn,10pt]{wlscirep}
\usepackage[utf8]{inputenc}
\usepackage[T1]{fontenc}
\usepackage{threeparttable}
\usepackage{float}
\usepackage{bm}
\usepackage{xr-hyper}
\usepackage{hyperref}
\title{Nodal error behind discrepancies between coupled cluster
   and diffusion Monte Carlo in hydrogen-bonded systems}

\author[1]{S. Lambie}
\author[1]{P. L\'opez R\'ios}
\author[1]{D. Kats}
\author[1,2,*]{A. Alavi}

\affil[1]{\textit{Max Planck Institute for Solid State Research, Heisenbergstra\ss e 1, 70569 Stuttgart, Germany}}
\affil[2]{\textit{Yusuf Hamied Department of Chemistry, University of Cambridge, Lensfield Road, Cambridge CB2 1EW, United Kingdom}}

\affil[*]{a.alavi@fkf.mpg.de}

\begin{abstract}
The small magnitude and long-range character of non-covalent
interactions pose a significant challenge for computational quantum
chemical and electronic-structure methods alike.
State-of-the-art coupled cluster (CC) theory and benchmark-grade
diffusion Monte Carlo (DMC) are ideally positioned to tackle these
problems, but concerning differences between both methods have been
reported in numerous studies of the interaction energy of
non-covalently bound dimers.
Given that the basic theoretical frameworks underpinning both methods
are exact in principle, the error must arise from one or several of
the approximations required to make the calculations computationally
tractable.
Here, we carry out a rigorous and systematic examination of the effect
of each of these approximations using the acetic acid dimer and water-peptide systems as convenient testing grounds.
Thanks to the use of stringently optimized backflow wave functions we
are able to find that the significant discrepancies are
dominated by the fixed-node error incurred
by the Slater-Jastrow DMC result, while errors in the CC calculations
do not significantly alter the result.
This finding, likely applicable to other hydrogen-bonded systems, helps
establish that CC should be regarded as the benchmark for these
systems, and can potentially guide the search for pragmatic solutions
to the fixed-node problem in the future.
\end{abstract}

\begin{document}

\flushbottom
\maketitle
% * <john.hammersley@gmail.com> 2015-02-09T12:07:31.197Z:
%
%  Click the title above to edit the author information and abstract
%
\thispagestyle{empty}

\section*{Introduction}

The quality and relevance of a computational model lie in its
ability to reproduce and predict experimentally observed phenomena.
However, obtaining reliable experimental data which can be directly
compared to computational results is a complicated and fraught
endeavor, especially for non-covalent interactions.
%
%One such domain in which experimental
%and theoretical results are, unfortunately, not directly comparable is the %strength of non bonded interactions between molecules.
%
In an experiment, the effects of temperature and solvent are embedded
in any measurement and must be disentangled from the measurement
itself; a non-trivial process which introduces errors that are
difficult to quantify.
By contrast, theoretical calculations establish the interaction energy
between two rigid molecules at $0$ K in the gas phase, with the comparison
to experiment being further exacerbated by the typically small
magnitude of these interactions.
High-quality quantum chemical calculations approximating the exact
full configuration interaction (FCI) result in the complete basis set
(CBS) limit provide crucial benchmark results to which the quality of
other theoretical models can be established.

%to accurately assess the performance of a quantum chemical method, datasets are used, for which high quality, systematically improvable reference data exists to benchmark against. 

%Benchmark datasets are vitally important for establishing high quality reference data, against which the performance and reliability of different methods can be determined. Such datasets exist for non-covalent interactions, largely due to the work of \v{R}ez\'a\v{c} and Hobza, who developed the A24,\cite{rezac:2013b} X40,\cite{rezac:2012} S22,\cite{jurecka:2006} S66,\cite{rezac:2011} and L7 datasets,\cite{sedlak:2013} and derivatives thereof to include non-equilibrium geometries.\cite{rezac:2011b, grafova:2010} These datasets report non-covalent interactions for increasingly large molecular systems typically using coupled cluster (CC) theory. 
	
In quantum chemistry, coupled cluster (CC) theory\cite{cizek:1966, paldus:1972} has long provided a systematically improvable, hierarchical approach for moving toward the exact solution of the Schr\"odinger equation of molecules of moderate
sizes as more excitations are included in the cluster operator. 
In turn, the diffusion quantum Monte Carlo (DMC) method has emerged
among electronic structure methods as one of the most accurate
approaches to solving the Schr\"odinger equation for systems with
several hundreds of electrons.
As computational resources have become more powerful over the past few
decades, CC has become applicable to larger system sizes while DMC has
seen improvements in the quality with which moderately-sized systems
can be described.
Comparisons of the two methods are, therefore, of great interest,
\cite{manten:2001, mella:2003, dubecky:2014, dubecky:2013} and several
recent studies in the literature have focused on disagreements found
between CC and DMC in weakly-interacting systems for which DMC is expected to
be particularly well-suited.  \cite{al-hamdani:2021,
villot:2022, shi:2025}

Both CC theory and the DMC method are, in principle, capable of exactly
solving the Schr\"odinger equation, \cite{witte:2015} but in practice
both require the use of a variety of approximations.
In the case of CC theory, the cluster operator is typically truncated
at the singles, doubles, and perturbative triples (CCSD(T)) level,\cite{raghavachari:1989} which provides a very good balance of accuracy and computational cost compared to CCSDT(Q) and CCSDTQ\cite{kallay:2005} for a majority of chemical properties.
\cite{rezac:2013, rezac:2013b}
However, the use of finite basis sets and the truncation of the
cluster operator introduces errors into the calculation that
grow as system size increases.
The DMC method requires controlled approximations, such as the use of
finite time steps and walker populations, and uncontrolled
approximations, such as the use of pseudopotentials and the
localization approximation they require.
However, the bulk of the error incurred by DMC is ultimately from the
fixed-node approximation, whereby the DMC energy depends on the
quality of the nodes of a trial wave function that guides the
stochastic sampling process.

\begin{figure}[H]
  \centering
    %\begin{tabular}{c}
 % \includegraphics*[width = 0.5\textwidth]{acoh_dimer_cc_results_alternative}\\
   \includegraphics*[width = 0.5\textwidth]{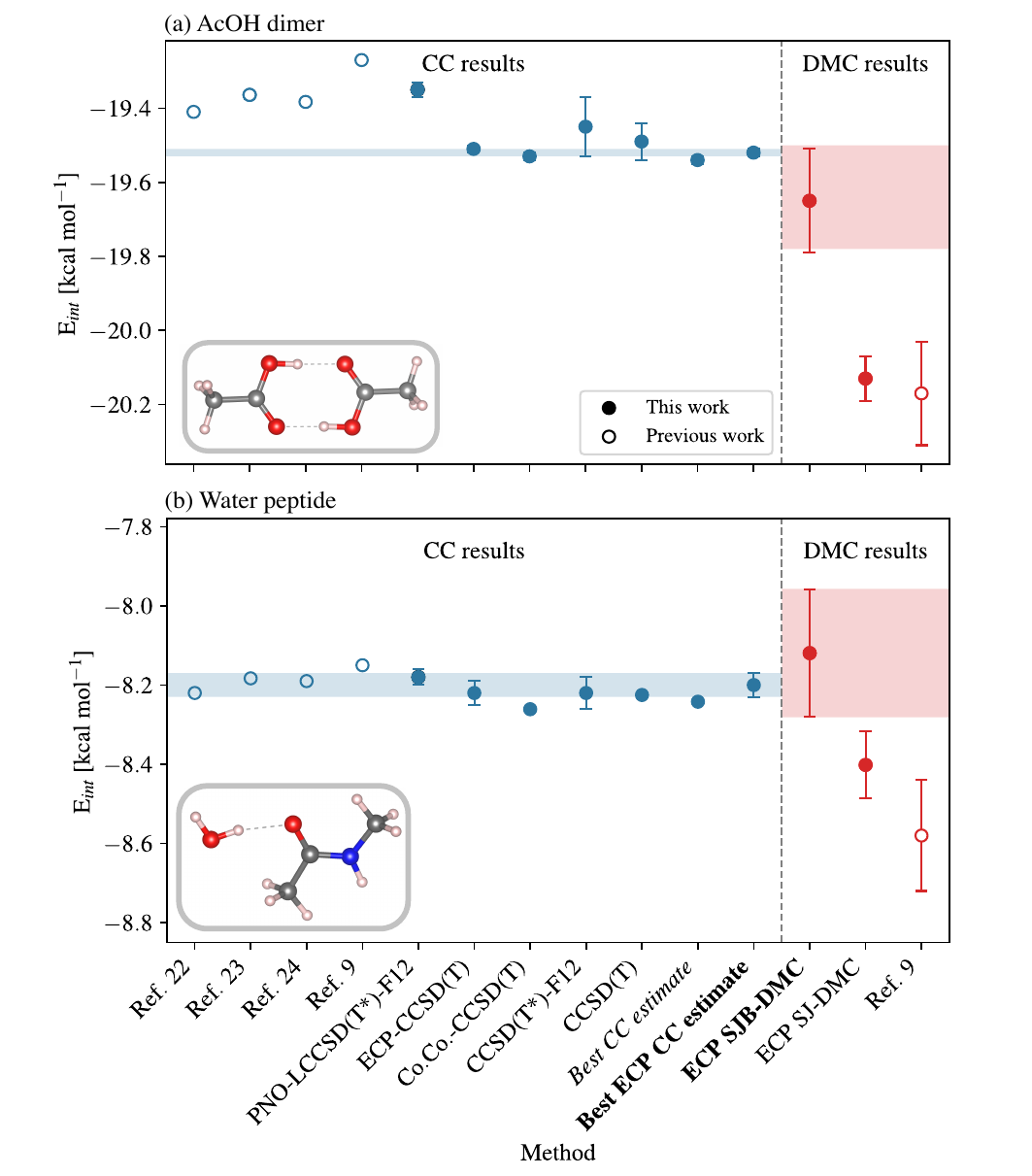}\\
   %\color{red} Alternative plot for presenting results\\
    %\includegraphics*[width = 0.5\textwidth]{acoh_dimer_cc_results_alternative}\\
    %    \end{tabular}
  \caption{
    Interaction energy of (a) the AcOH dimer and (b) the water-peptide dimer obtained using CC theory (blue) and DMC (red).
    The interaction energies that allow for the most direct comparison between each methodology (bold)
    with a `two-sigma' confidence interval are shown as a shaded
    area.
    For the CC calculations, 0.5CP-corrected values are reported
    with non-statistical error bars extending from the CP-uncorrected
    to the CP-corrected result; note that reference CC values are
    plotted as single points.
    The error bars on the DMC values are purely statistical in
    nature, representing 95\% (two-sigma) confidence intervals, and
    include optimization uncertainty where pertinent, see
    text.
    The insets show the structures considered with C atoms in grey,
    O atoms in red, H atoms in pink and N atoms in blue.}
  \label{fig:results}
\end{figure}

The finding of significant discrepancies between the two methods for
non-covalent interaction energies \cite{al-hamdani:2021, villot:2022}
stimulated a flurry of additional studies investigating the possible
sources of disagreement. \cite{nakano:2024, nakano:2024b, benali:2020,
gray:2024, schafer:2024, lambie:2024, lao:2024}
The issue was initially thought to be restricted to large,
dispersion-bound systems, but substantial deviations were also found
for small H-bound systems.  \cite{shi:2025}
To date, the origin of the discrepancy between the two state-of-the-art
methods has not been definitively identified.
Here, we focus on the two case-study examples of the 16-atom acetic
acid (AcOH) dimer and 15-atom water-peptide system, corresponding to
systems 20 and 4 of the hydrogen-bonded portion of the S66 set,
respectively;\cite{rezac:2011} see insets in Fig.\ \ref{fig:results}.
For the AcOH dimer, a discrepancy of
about $0.8$ kcal mol$^{-1}$ has been reported between the DMC value of
$-20.17(7)$ kcal mol$^{-1}$ and canonical CC value of $-19.39(2)$ kcal
mol$^{-1}$,\cite{shi:2025} while for the water-peptide system, the difference between the DMC result of $-8.58(7)$ kcal mol$^{-1}$ and the CC result of $-8.20(2)$ kcal mol$^{-1}$ is about 0.4 kcal mol$^{-1}$.\cite{shi:2025}
The AcOH dimer and water-peptide systems thus provide a wonderful opportunity for probing
sources of error between the two methods thanks to the large energy
discrepancies and small system sizes.

\section{Coupled cluster results}

\begin{table}
  \centering 
  \caption{
    Interaction energy of the AcOH dimer from CC theory in kcal
    mol$^{-1}$, expressed as the 0.5CP-corrected value plus/minus
    half of the total CP correction except where otherwise stated.
    A bold typeface is used for our best CC theory AcOH dimer
    interaction energy with ECPs.}
    %
 %   $Footnotes define the abbreviations used. }
  \label{acoh_numbers}
 % \begin{threeparttable}[b]
    \begin{tabular}{llr@{.}lr@{.}l} 
    \hline
    \hline
    \multicolumn{6}{c}{This work} \\
    \hline
    \hline  
    In text-referral & Treatment & \multicolumn{4}{c}{E$_{\text{int}}$ [kcal mol$^{-1}$]} \\
    & &   \multicolumn{2}{c}{(AcOH)$_{2}$}  &   \multicolumn{2}{c}{Water-peptide} \\
    \hline
    %FP-CCSD(T)       &
    %   HF/DF-aV5Z+MP2/DF-aV\{Q,5\}Z + $\Delta$CCSD(T)/aV\{T,Q\}Z   &
    %   $-19.41(5)$  \\
    CCSD(T)        &
       DF-HF/aV5Z + CCSD(T)/aV\{T,Q\}Z &
       $-19$&$49(5)$  & $-8$&$225(5)$\\
    CCSD(T*)-F12      &
       DF-HF-F12/aV5Z + CCSD(T*)-F12/aVQZ &
       $-19$&$45(8) $  & $-8$&$22(4)$\\
    %HA-CCSD(T)        &
    %   HF/aVQZ+ CCSD(T)/haV\{T,Q\}Z &
    %   $-19$&$429(4)$  \\
    \hline
    Co.Co.-CCSD(T)    &
       DF-HF/aCV5Z + CCSD(T)/aCV\{T,Q\}Z &
       $-19$&$53(1)$  & $-8$&$261(3)$\\
    ECP-CCSD(T)       &
       DF-HF/aV5Z-eCEPP + CCSD(T)/aV\{T,Q\}Z-eCEPP &
       $-19$&$51(1)$  & $-8$&$22(3)$\\
    \hline
    PNO-LCCSD(T*)-F12 &
       DF-HF-CABS/aV5Z + PNO-LCCSD(T*)-F12/aV5Z &
       $-19$&$35(2) $  & $-8$&$18(2)$\\
    \hline
    \textit{Best CC estimate}	  &
       HF/aCV5Z + CCSD(T)/aCV\{T,Q\}Z +  \\
       &$\Delta$SVD-DC-CCSDT+/aVTZ &
       $\textit{{-19}}$&${\textit{54(1)}}$ & $\textit{-8}$&$\textit{242(3)}$ \\
       \textbf{Best ECP CC estimate} & DF-HF/aV5Z-eCEPP + CCSD(T)/aV\{T,Q\}Z-eCEPP+ \\
       &$\Delta$SVD-DC-CCSDT+/aVTZ & $\bm{-19}$&$\bm{52(1)}$ & $\bm{-8}$&$\bm{20(3)}$ \\
    \hline
    \hline
    \multicolumn{6}{c}{Reference values} \\
    \hline
    Ref. \citenum{rezac:2011b}     &
       HF/aV\{T,Q\}Z + MP2/aV\{T,Q\}Z + $\Delta$CCSD(T)/haV\{D,T\}Z (CP)&
       $-19$&$41$ & $-8$&$22$ \\
    Ref. \citenum{kesharwani:2017} &
       HF-CABS/V5Z-F12 + MP2-F12/aV\{T,Q\}Z-F12 +   \\
     &  $\Delta$CCSD(F12*)/VTZ-F12 +$\Delta$(T)/haV\{D,T\}Z&  $-19$&$364(5)$ & $-8$&$18(1)$\\
    Ref. \citenum{nagy:2022}       &
       HF-CABS/aVQZ-F12+ MP2/aV\{T,Q\}Z-F12 +  \\
     & $\Delta$CCSD(*)/haV\{D,T\}Z+$\Delta$(T)/haV\{T,Q\}Z (mixed)&
  $-19$&$38$ & $-8$&$19$ \\
    Ref. \citenum{shi:2025} &
       Ref. \citenum{kesharwani:2017} + CCSD(cT)-fit (mixed) &
       $-19$&$27$ & $-8$&$15$ \\
    \hline
    \hline
    \end{tabular}
   % \begin{tablenotes}
   %   \footnotesize
      %\item[g] CCSD(F12*)-MP2-F12/cc-pVQZ-F12(0.5CP) +
      %   (T)/haV\%{D,T\}Z(0.5CP). 
      %\item[h] HF-CABS/aVQZ-F12(0.5CP)+ MP2/aV\{T,Q\}Z-F12(CP) +
      %   CCSD(*)-MP2/haV\{D,T\}Z(CP)+ (T)/haV\{T,Q\}Z(CP). 
      %\item[i] CCSD(cT)
    %  \item[]
    %     FC:\ frozen core;
     %    CP:\ counter-poise;
      %   Co.Co.:\ core correlated;
      %  CABS:\ complementary auxiliary basis set. 
    %\end{tablenotes}
 % \end{threeparttable}
\end{table}

%\begin{figure}[H]%
%	\centering
 %   \begin{tabular}{c}
	%\includegraphics*[width = 0.5\textwidth]{acoh_dimer_cc_results_edited2}\\
%   \color{red} Alternative plot for presenting results\\
%    \includegraphics*[width = 0.5\textwidth]{acoh_dimer_cc_results_alternative}\\
 %       \end{tabular}
  %  \caption{Coupled cluster results for the AcOH dimer using a variety of different approximations. Our best estimate counterpoise (CP) and non-counterpoise (raw)  interaction energies are shaded in grey (upper plot)/blue in lower plot.}

In CC theory, `gold standard' results are obtained by truncating the
CC operator at the CCSD(T) level, which provides an excellent balance
of accuracy and computational cost due to a systematic cancellation of
errors. \cite{helgaker:2004, rezac:2013b, rezac:2013} 
However, a beyond-`gold' standard is required for
benchmarking quantum chemical methods, taking into consideration all
possible sources of error in CCSD(T). 
We have studied the convergence of CC results with respect to the
basis set (section \ref{basis_sets}), electronic excitations (section
\ref{core}), local implementation (section \ref{local_cc}), and CC
operator (section \ref{pt}). 

\subsection{Approaching the complete basis set limit}\label{basis_sets}

%In addition to the truncation of the CC operator, results need to be
%converged with the size of the basis set.
%

%Furthermore, augmentation functions are important for accurately capturing intermolecular energies.\cite{rezac:2016} 
	
%Despite the large size and diffuse nature of the basis sets required for intermolecular interactions, 

All basis sets are finite and, as a result, incomplete, meaning that
both basis set superposition error (BSSE) \cite{liu:1973} and basis
set incompleteness error (BSIE) \cite{davidson:1986} require careful
management.
The counterpoise (CP) correction is the  most common approach for
dealing with BSSE, \cite{boys:1970} which allows the CBS limit to be
approached more smoothly and quickly than with uncorrected (raw)
values. \cite{halkier:1999, burns:2014} 
Raw and CP-corrected results typically provide a lower and upper
bound, respectively, for the CBS value, within a specific CC operator
truncation, \cite{burns:2014} and for this reason we report our CC
results as the arithmetic mean of the raw and CP-corrected values
plus/minus a half of the full CP correction, so that the
`two-sigma' interval, which corresponds to the 95\% confidence
interval for statistical standard errors, covers the full CP range.
This simplifies comparison with DMC where error bars occur naturally,
but bear in mind that our CC `error bars' are not statistical in
nature.
%
%CP corrected results generally converge more quickly and smoothly to the CBS limit than the raw calculations.\cite{halkier:1999, burns:2014} 
%
In the CBS limit, both the BSIE and BSSE vanish, along with the need
for CP corrections.

Correlation energy is slowly convergent with basis-set size,
\cite{hobza:1996, shaw:2018, helgaker:1997, steele:2009,
kodrycka:2019} and basis sets of at least augmented triple zeta (aVTZ) size
are essential for CC results to be considered reliable for
non-covalent interactions.\cite{rezac:2016}
For this reason, the largest computationally-affordable basis sets are
used in conjunction with techniques to approximate the CBS limit, such
as focal-point (FP) approaches,\cite{burns:2014, marshall:2011} CBS
extrapolations\cite{halkier:1999, helgaker:1997} and explicitly
correlated F12 methods.\cite{knizia:2009}

FP approaches \cite{burns:2014, marshall:2011} exploit the fact that
the total energy of a post-Hartree-Fock method can be decomposed into
the Hartree-Fock (HF) component and various correlation energy
contributions calculated using basis sets of different sizes, taking
advantage of the fact that the energy offset between methods is fairly
consistent.
The FP approach can then be coupled further with a CBS extrapolation,\cite{halkier:1999, helgaker:1997}
so that the correlation energy component of the calculation can be
obtained in smaller basis sets and extrapolated to the CBS limit. 
Explicitly correlated methods significantly improve basis set
convergence by including geminal functions that rely on the
inter-electronic distance into the wavefunction
ansatz. \cite{adler:2007, knizia:2009}
The (T) treatment does not contain any F12 terms\cite{patkowski:2013}
and, therefore, the (T) energy must be scaled (see Section
\ref{methods}) resulting in the CCSD(T*)-F12 method.
Performance of CCSD(T*)-F12 with aVTZ and aVQZ basis sets is known to be comparable to CBS extrapolated canonical CCSD(T)/aV\{Q,5\}Z results for non-covalent interactions.\cite{sirianni:2017} 
Here, the CCSD(T*)-F12 results agree with the CCSD(T) results
(Table \ref{acoh_numbers}), to which it can be most directly compared,
determining that both the explicitly correlated method and CBS
extrapolated results converge toward the CBS limit for the interaction
energy of the AcOH dimer and water-peptide system at the CCSD(T) level of theory. 
For the AcOH dimer, we additionally test the truncated heavy-atom only augmented\cite{marshall:2010} and density fitted\cite{weigend:2002, weigend:2002b} basis sets with minimal changes to the results, see Section S1 of the supplementary information.

%Previous calculations of the interaction energy for the AcOH dimer
%have reported values of $-19.41$, \cite{rezac:2011b} $-19.36$,
%\cite{kesharwani:2017} and $-19.38$ \cite{nagy:2022} kcal mol$^{-1}$.
%
In general, we find that our CC results for the AcOH dimer and the water-peptide system are in good agreement
with the literature (Table \ref{acoh_numbers}).\cite{rezac:2011b, kesharwani:2017, nagy:2022} However, the results of the present study can be expected to be more accurate
than previously reported values thanks, in part, to our use of larger basis
sets. Additionally, the previously reported values use the FP approach extensively -- that is, they make two key assumptions that could introduce an undefined error into the results. Firstly, all reference values assume that the difference in energy between lower order methods, such as MP2 and CCSD(T) is constant and unchanging as the size of the basis set is increased. Secondly, they assume that these energetic differences are converged in small basis sets, such as the heavy-atom augmented double-triple-zeta extrapolation $\Delta$CCSD(T) energetic correction, used in Ref \citenum{rezac:2011b}. While the FP approach is generally applicable, and introduces small errors, these errors are not necessarily systematic\cite{burns:2014} and are relevant on the scale of interest for the non-covalent interactions reported in the present study. As such, it is imperative to control all sources of error in the coupled cluster calculations and by calculating the correlation energy in large basis sets with minimal usage of the FP approach, we control these additional sources of error in the present calculations. 
It should also be noted that some of the previous studies referenced here
reported individual interaction energies containing a specific amount
of CP correction, and are therefore comparable with the outer limits
of our `two-sigma' confidence intervals. 

\subsection{Core electron treatments}\label{core}

The frozen core approximation is routinely used in CC calculations,
but its effect on the results is rarely considered. In the present study, calculations invoke the frozen core approximation unless otherwise stated.
Alternative treatments of the core region exist, in particular, the
explicit inclusion of core correlation in the calculation (Co.-Co.-CCSD(T)) through the
use of correlation consistent core-valence basis sets, aCVX Z, X = T, Q, where the C indicates that the basis set is suitable for correlating core electrons,\cite{peterson:2002} or the core can be treated
using effective core potentials (ECP-CCSD(T)), as commonly used in DMC
calculations; specifically, we use energy-consistent correlated
electron pseudopotentials (eCEPPs). \cite{trail:2017}

We have performed calculations of the AcOH dimer and water-peptide
system with all three treatments of the core, and we find that a very
small stabilisation on the order of 0.04 kcal mol$^{-1}$ arises as a
result of including core correlation in the calculation. 
Therefore, we conclude that core treatment is unlikely to be
responsible for the significant discrepancy between the CC and DMC
results. 

\subsection{Local coupled cluster}\label{local_cc}

Local methods \cite{riplinger:2013, nagy:2018, ma:2018b} are designed
to reach larger system sizes with high accuracy and are often used to
obtain CCSD(T)/CBS reference values for non-covalent interaction
energies of large molecules. \cite{al-hamdani:2021, nagy:2024,
hansen:2025, lao:2024, villot:2022}
Here, we focus on the most accurate of the local methods,
PNO-LCCSD(T*)-F12, \cite{ma:2018b} where the combination of local and
scaled F12 methods is highly effective at reducing the basis-set and
domain errors. \cite{ma:2018b, jakubikova:2006, hill:2006,
krause:2012}
However, both basis set and local errors increase as the system size grows,\cite{ma:2018b, altun:2021} and, therefore, the fidelity of local methods to the canonical result is expected to deteriorate with increasing system size. \cite{ma:2018b, hansen:2025}
%
%Many different flavors of local approximation exist, including the domain
%local pair natural orbital (DLPNO),\cite{riplinger:2013} local natural orbital\cite{nagy:2018} and the pair natural orbital method
%in conjunction with (T*) scaled explicit correlation (PNO-LCCSD(T*)-F12). \cite{ma:2018b}
%
%Importantly, different implementations of local methods agree with one another, giving confidence to the individual implementations for capturing large, molecular dispersive interactions. \cite{villot:2022} 
While the reduced scaling offered by local methods is not required to
simulate the systems considered in this study due to their small size,
we test this method for its performance compared to the canonical
result to better understand possible sources of error.

%

%
%However, the PNO-LCCSD(T*)-F12 method has been benchmarked
%against canonical CCSD(T) results for the S22 dataset and found to agree to within 0.5 kcal mol$^{-1}$. \cite{ma:2018b, hansen:2025} 
%
The PNO-LCCSD(T*)-F12 calculations are a small underestimation of the CCSD(T*)-F12 results (Table \ref{acoh_numbers}), as expected by the method developers.\cite{ma:2018, ma:2018b}
While the low basis-set error of local methods\cite{hampel:1996,
saebo:1993, runeberg:1999, schutz:1998} would appear to indicate a low
error on local CC values, we advise that confidence intervals be
estimated conservatively for local calculations, extending beyond
basis-set error, until the extrapolation of local results to the
canonical limit is well-established. \cite{sorathia:2020}

%Importantly, we show that the local CC result has a small BSSE,\cite{hampel:1996, saebo:1993, runeberg:1999, schutz:1998} however this does not necessarily imply direct agreement with the canonical result. While the local method does perform very well, being only 0.05 kcal mol$^{-1}$ different from the canonical result, it is ultimately an approximation of the canonical CCSD(T) result.
%

\subsection{Beyond gold standard coupled cluster}\label{pt}

For molecular dimers it has been found that CCSD(T) regularly
outperforms CCSDT, \cite{rezac:2013, rezac:2013b, demovicova:2014,
lambie:2024, semidalas:2025} and that CCSDT(Q) and CCSDTQ typically
produce results that are converged with respect to the cluster
operator for non-covalent interactions. \cite{rezac:2013}
%Up to this point, we have discussed systematic improvability within
%canonical CC with respect to the cluster operator and how to approach the %CBS limit.
%%
However, the perturbative treatment of excitations in CC theory causes
diverging results in the metallic limit.\cite{schafer:2024,
benali:2020, shepherd:2013}
%However, as the size of polyaromatic hydrocarbon molecules increases toward the infinite limit, gap between the highest occupied molecular (HOMO) -lowest unoccupied molecular orbital (LUMO) (bandgap in extended systems) diminishes toward the metallic limit,\cite{shen:2018, tonshoff:2020, novoselov:2004} and the perturbative treatment of (T) in CCSD(T), causes diverging results.\cite{schafer:2024, benali:2020, shepherd:2013}
%
Since the discrepancy between DMC and CC results was first reported
for increasingly large dispersion-bound C-based molecules
\cite{al-hamdani:2021, villot:2022, shi:2025} with diminishing
HOMO-LUMO gaps, \cite{shen:2018, tonshoff:2020, novoselov:2004} lack
of convergence with respect to the truncation of the CC operator must
be considered as a possible source of error. 

Even for the small systems considered in this study, the \textit{N}$^{9}$ scaling of CCSDT(Q) means
that canonical calculations in large basis sets are unattainable.
Therefore, post-CCSD(T) corrections can be obtained by approximating
the CC operator through the use of CCSD(cT) \cite{schafer:2024} or rank-reduced distinguishable cluster (SVD-DC-CCSDT+)\cite{rickert:2025, kats:2014, kats:2013, kats:2019} methods, or by using very small,
truncated basis sets to carry out canonical CCSDT(Q)
calculations. \cite{semidalas:2025}

Each of these approaches has its own limitations. 
CCSD(cT) \cite{schafer:2024} approximates CCSDT, which is known to be
somewhat inaccurate for non-covalent interactions.  \cite{rezac:2013,
lambie:2024, rezac:2013b, demovicova:2014, semidalas:2025}
SVD-DC-CCSDT+ \cite{rickert:2025} enables larger basis sets to be used
but is a new method that has currently only been benchmarked to the
A24 dataset.\cite{rezac:2015, lambie:2025} Additional benchmarking of the SVD-DC-CCSDT+ method for the AcOH dimer and water-peptide system is shown in the supplementary information.
Canonical CCSDT(Q) results \cite{semidalas:2025} require the use of
small basis sets, implying that these corrections cannot be expected
to be converged with respect to basis set size. \cite{smith:2014}
Nevertheless, these three approaches result in small post-CCSD(T)
corrections of, for the AcOH dimer $0.12$ kcal mol$^{-1}$, $-0.01$ kcal mol$^{-1}$ and
$0.055$ kcal mol$^{-1}$, and for the water-peptide system of 0.05 kcal mol$^{-1}$, 0.019 kcal mol$^{-1}$ and 0.012 kcal mol$^{-1}$, respectively, for
fitted CCSD(cT),\cite{shi:2025} SVD-DC-CCSDT+/aVTZ, and
CCSDT(Q)/VDZ(d,s).\cite{semidalas:2025}
Interestingly, all of the results considering higher order excitations in the cluster operator than CCSD(T) have either
negligible effects or reduce the magnitude of the CC result, thereby
increasing the discrepancy between the DMC and CC results.
We, therefore, conclude that the discrepancy with DMC is highly unlikely
to be sourced from truncation of the CC operator.

\subsection{Other considerations and the best coupled cluster estimate}

Our best CC estimates for the interaction energy of the AcOH dimer and
water-peptide system using CC-based methodologies are $-19.54(1)$
kcal mol$^{-1}$ and $-8.242(3)$ kcal mol$^{-1}$, respectively.
These calculations improve upon previously reported values by using
larger basis sets for the correlation energy component of the
calculations, considering excitations from the deep core and an
approximate post CCSD(T) correction.
Other possible sources of error in our best estimate could arise from
scalar relativistic effects and diagonal Born-Oppenheimer corrections,
however, these are known to be negligibly small for light atoms,
\cite{karton:2021} such as C, N, O, and H found in (AcOH)$_2$ and the
water-peptide system, and are also absent from the DMC calculations.

Since none of the approximations employed in CC theory are able to
explain the discrepancy with the DMC results, we turn to examining
possible sources of error arising from the approximations involved in
the DMC calculations.
To directly compare between CC and DMC, we take the ECP-CCSD(T) result
and add a post-CCSD(T) correction to obtain our `best ECP CC
estimate.' 

\section{Diffusion quantum Monte Carlo results}

\subsection{Slater-Jastrow results}

We have run DMC calculations of the AcOH, (AcOH)$_2$, water, peptide,
and water-peptide systems using the Slater-Jastrow (SJ) trial wave
function and ECPs to represent ionic cores, namely eCEPPs,
\cite{trail:2017} with which we recover ECP SJ-DMC results within
uncertainty of those reported in Ref.\ \citenum{shi:2025}, see Table
\ref{table:dmc} and Fig.\ \ref{fig:results}.

\begin{table}
  \centering 
  \caption{
    \label{table:dmc}
    DMC interaction energy of the AcOH dimer and the water-peptide
    system, in kcal mol$^{-1}$. Uncertainties shown represent 95\%
    (2-sigma) confidence intervals. A bold typeface is used for our
    best DMC interaction energy.
  }
  \begin{tabular}{lr@{.}lr@{.}l} 
    \hline
    \hline  
    \multicolumn{5}{c}{This work} \\
    \hline
    \hline  
    Method
    & \multicolumn{4}{c}{E$_{\text{int}}$ [kcal mol$^{-1}$]} \\
    & \multicolumn{2}{c}{(AcOH)$_{2}$}
    & \multicolumn{2}{c}{Water-peptide} \\
    \hline
    AE SJ-DMC
      & $-20$&$29(16)$           & \multicolumn{2}{c}{-} \\
    ECP SJ-DMC
      & $-20$&$13(6)$            & $-8$&$40(8)$ \\
    \textbf{ECP SJB-DMC}
      & $\bm{-19}$&$\bm{65(14)}$ & $\bm{-8}$&$\bm{12(16)}$ \\
    \hline
    \hline
    \multicolumn{5}{c}{Reference values} \\
    \hline
    ECP SJ-DMC \cite{shi:2025}
      & $-20$&$17(14)$            & $-8$&$58(14)$ \\
    \hline
    \hline
  \end{tabular}
\end{table}

There are various potential sources of error in the DMC results.
DMC calculations are performed at finite time steps $\tau$, and the
resulting energies must be extrapolated to zero time step as we report
in Section S2 the supplementary information.
Population control bias is typically negligible and can be removed by
extrapolation along with time-step bias, \cite{needs:2020} which we
have done.
We have verified that the error incurred by the use of ECPs is
negligible on the scale of interest by computing the all-electron (AE)
SJ-DMC interaction energy of the AcOH dimer, reported in Table
\ref{table:dmc}, which agrees within uncertainty with its ECP
counterpart.
We have tested using distinct Jastrow parameters for
symmetry-inequivalent atoms, \cite{dubecky:2019} but we find that this
does not change the interaction energy estimate for (AcOH)$_2$.
The source of the remaining error in DMC is the mismatch between the
nodes of the trial wave function and those of the exact wave function,
which is referred to as the fixed-node error.
Note that BSIE can be regarded as part of the fixed-node error in DMC;
our use of the relatively large aug-cc-pVTZ basis set should provide a
good starting point for this source of fixed-node error.

\subsection{Beyond Slater-Jastrow nodes: backflow results}

The use of multideterminant expansions is a popular approach for
obtaining beyond-SJ nodes to study electronic excitations of small
molecules, \cite{filippi:1996, brown:2007, seth:2011, petruzielo:2012,
morales:2012, giner:2016, scemama:2018, scemama:2019, dash:2018,
dash:2019, dash:2021} but it suffers from significant size-consistency
issues that preclude the calculation of accurate interaction energies,
especially for the system sizes of interest here, so we have instead
chosen to apply backflow \cite{feynman:1956, lee:1981, kwon:1998,
holzmann:2003, lopezrios:2006} to $\Psi_\text{S}$.
In the Slater-Jastrow-backflow (SJB) wave function the arguments of
the single-particle orbitals are replaced with quasiparticle
coordinates that depend on the positions of all other electrons,
smoothly altering the shape of the SJ nodal surface; see Section S2
of the supplementary information for further details.
SJB-DMC is often capable of removing about half of the fixed-node
error in the SJ-DMC energy.  \cite{brown:2007, seth:2011, needs:2020}
The use of backflow incurs a significant increase in the computational
cost of DMC calculations, so it is typically not used for large
systems, but the expense is manageable for the systems studied here
when ECPs are used.
The topical, very flexible neural-network wave functions
\cite{pfau:2020, hermann:2020} incorporate Jastrow-, backflow-,
orbital-, and multideterminant-like degrees of freedom, but we have
opted against using these because they incur potentially greater
costs, require specialized workflows and hardware, and preclude
straightforward comparisons with SJ-DMC.

VMC-based optimization of Jastrow factors is usually performed without
considering the presence of statistical uncertainties on the resulting
parameters arising from the stochastic nature of the optimization
process.
This is because the Jastrow factor parameters in the SJ wave function
do not affect the value of (all-electron) SJ-DMC energies at $\tau=0$,
but there are scenarios where wave function parameters
significantly affect expectation values, \cite{haupt:2023, filip:2025,
spink:2016} as is the case of the DMC energy in the presence of
backflow.
Ignoring optimization uncertainty can thus result in the perception
that SJB-DMC energies behave erratically and are inconsistent across
systems.

The optimization uncertainty on the SJB-DMC energy $\sigma_\text{opt}$
depends on the number of real-space configurations $n_\text{opt}$ used
in VMC-based correlated-sampling optimization via $\sigma_\text{opt}^2
= \alpha / n_\text{opt}$, where $\alpha$ is a system-dependent
unknown, at sufficiently large $n_\text{opt}$, see section S2 of the
supplementary information.
For a given $n_\text{opt}$, the optimization uncertainty on the
SJB-DMC energy can be evaluated by running multiple independent random
instances of optimization at that sample size, each followed by a
SJB-DMC run.
It is therefore possible to perform this procedure at a trial value of
$n_\text{opt}$ to obtain $\sigma_\text{opt}$ in order to determine
$\alpha$, and thus be able to identify the value of $n_\text{opt}$
necessary to obtain any given target accuracy

\begin{figure*}
  \centering
  \includegraphics[width=0.48\textwidth]{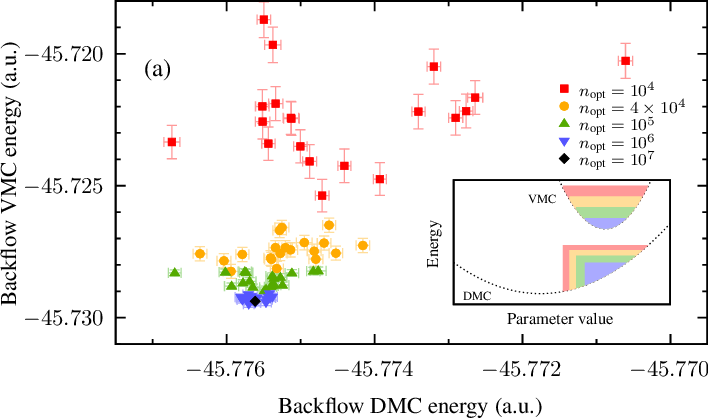}
  ~
  \includegraphics[width=0.48\textwidth]{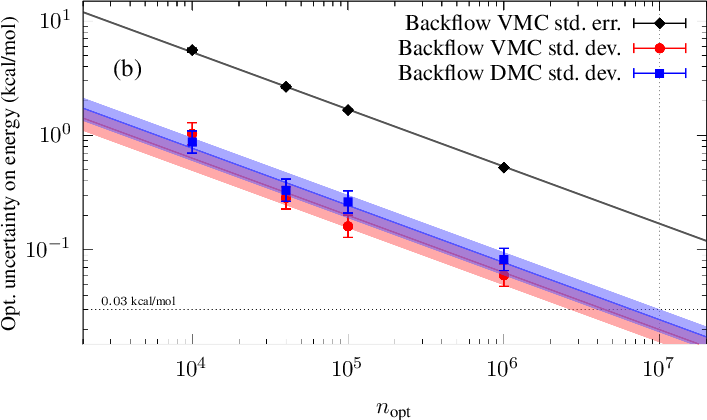}
  \caption{
    (a) Scatter plot of the SJB-VMC energy as a function of SJB-DMC
    energy of the ECP AcOH monomer for 20
    independent random instances of wave function optimization at each
    of four optimization sample sizes $n_\text{opt}$, and for the
    final optimization performed at $n_\text{opt}=10^7$;
    note the lack of visible correlation between VMC and DMC energies
    and the evident mismatch of the location of the VMC and DMC
    energy minima, schematically represented in the inset.
    (b) Resulting uncertainty (1-sigma confidence interval) on the
    SJB-VMC and SJB-DMC energy arising from the stochastic nature of
    optimization as a function of $n_{\rm opt}$, computed as the
    standard deviation of the corresponding energies in
    Fig.\ \ref{fig:opt_figure}(a), in log-log scale.
    The target optimization uncertainty and chosen sample size are
    shown as dotted lines, and solid lines are linear fits in log-log
    scale to the data points using a fixed slope of $-1/2$.
    Shaded areas and errorbars represent 95\% (two-sigma) confidence
    intervals.
    Also shown is the (average) standard error on the mean value of
    the $n_\text{opt}$ local energies used in optimization, which
    is an order of magnitude greater than the optimization uncertainty
    on the backflow VMC energy:\@ correlated-sampling optimizers can
    determine the location of the energy minimum to much better
    accuracy than the statistical resolution of the local energy
    sample mean would suggest.
    \label{fig:opt_figure}
  }
\end{figure*}

To verify that the relationship between $\sigma_\text{opt}$ and
$n_\text{opt}$ is applicable in practice, we have computed the
optimization uncertainty on the SJB-DMC energy of the AcOH monomer,
$\sigma_\text{opt}^\text{AcOH}$, for several values of $n_\text{opt}$.
In Fig.\ \ref{fig:opt_figure}(a) the energies from individual
instances of optimization at each sample size are shown, and Fig.\
\ref{fig:opt_figure}(b) is a log-log plot of
$\sigma_\text{opt}^\text{AcOH}$ as a function $n_\text{opt}$, which
confirms that the optimization uncertainty remains proportional to
$n_\text{opt}^{-1/2}$ in the entire range of $n_\text{opt}$ tested.

Using the average value of $\alpha$ obtained from these data, we pick
a final optimization sample size of
$n_\text{opt}^\text{AcOH}=n_\text{opt}^\text{(AcOH)$_2$}=10^7$ to
achieve a target optimization uncertainty on the interaction energy of
$\sigma_\text{opt}=0.071$ kcal mol$^{-1}$, set so as to obtain a total
uncertainty on the interaction energy below 0.10 kcal mol$^{-1}$, see
Fig.\ \ref{fig:opt_figure}(b); note that
$\sigma_\text{opt}^\text{AcOH} = \sigma_\text{opt} / \sqrt 6 \approx
0.029$ kcal mol$^{-1}$, as discussed in section S2 of the
supplementary information.

For the heterogeneous water-peptide system we choose to split the
target optimization uncertainty of $\sigma_\text{opt}=0.071$ kcal
mol$^{-1}$ in proportion to the square root of the number of electrons
in each of the systems, to keep it in line with the stochastic
uncertainty, and thus arrive at
$\sigma_\text{opt}^\text{water}=0.023$ kcal mol$^{-1}$,
$\sigma_\text{opt}^\text{peptide}=0.045$ kcal mol$^{-1}$, and
$\sigma_\text{opt}^\text{w.-p.}=0.050$ kcal mol$^{-1}$.
In this case, we evaluate the optimization uncertainty at a single
value of $n_\text{opt} = 3 \times 10^4$, and we thus determine target
optimization sample sizes of $n_\text{opt}^\text{water} = 1.1\times
10^7$, $n_\text{opt}^\text{peptide} = 10^6$ and
$n_\text{opt}^\text{w.-p.} = 7.5 \times 10^6$.

Our final SJB-DMC interaction energies of $-19.65(14)$ kcal mol$^{-1}$
for (AcOH)$_2$ and $-8.12(16)$ kcal mol$^{-1}$ for water-peptide,
given in Table \ref{table:dmc} and plotted in Fig.\ \ref{fig:results},
are in excellent agreement with our best ECP CC energy estimates of
$-19.52(1)$ kcal mol$^{-1}$ and $-8.20(3)$ kcal mol$^{-1}$,
respectively, strongly supporting the hypothesis that the main cause
of the disagreement between CC and SJ-DMC results in H-bonded systems
is the fixed-node
error.

\section*{Discussion}

We have considered various potential sources of error in CC and DMC
calculations of the AcOH dimer and water-peptide system, and find that
the lifting of various approximations does not significantly alter
the CCSD(T) result. 

%while we are able to find a modest increase of the CC interaction
%energy of about $0.2$ kcal mol$^{-1}$ with respect to the previously
%reported canonical result, 
In both of the case-study examples considered in this work, the use of
backflow brings the magnitude of the DMC interaction energy towards
the CC result, in such way that the two methods are now in concordance
with one another.
From this we come to the conclusion that the fixed-node error in the
SJ-DMC calculations is responsible for the vast majority of the
disagreement between the methods reported in the literature for this
class of systems.
This implies that the inconsistencies in the nodes of the HF wave
function (which the SJ wave function inherits) between monomer and
dimer, while arguably small given the weakness of the interaction,
suffice to put DMC at a disadvantage with respect to CC.
The application of backflow will undoubtedly be crucial in resolving
the discrepancies between CC and DMC in dispersion-bound systems too.

In order to obtain meaningful backflow results we have given explicit
consideration to the uncertainty introduced by the stochastic nature
of the optimization process, which accounts for the largest share of
the uncertainty on our final result.
Establishing the magnitude of this optimization uncertainty is crucial
in ensuring that DMC energies obtained using trial wave functions with
stochastically-optimized nodes are well-defined quantities not
affected by uncontrolled noise.

Using backflow in the way we have in our present work is not
necessarily a practical or affordable solution to the fixed-node error
problem of SJ-DMC.
Our SJB-DMC calculations have used 2.9 million core-hours on an AMD
EPYC cluster to obtain the interaction energy of (AcOH)$_2$, and
1.5 million for the water-peptide system, while their SJ-DMC
counterparts cost 0.42 and 0.27 million core-hours, respectively.
On top of this, we have used an unrefined approach with conservative
choices of run parameters to establish the optimal value of
$n_\text{opt}$, adding 1.2 and 3.4 million core-hours to the cost of
the (AcOH)$_2$ and water-peptide calculations, respectively.
Despite its costliness, the backflow ansatz is trivially size
consistent, unlike multideterminant expansions, and the optimization
protocol proposed in the present work prevents this property from
being adversely affected by the stochastic optimization process.
This approach allows backflow to be deployed in cases where beyond-SJ
nodes are important, prompting further work to make the use and
optimization of backflow more affordable in practice.

\section{Computational details}\label{methods}

In keeping with previous works, we utilize the S66
geometries\cite{rezac:2011} for the AcOH dimer (system 20) and the
water-peptide system (system 4) throughout the study, as available
from the Benchmark Energy and Geometry Database, \cite{rezac:2008} see
Section S4 of the supplementary information.

When calculating the interaction energy, no relaxation effect of the monomers is taken into account. Intermolecular energies are calculated with the super-molecular approach; for the AcOH dimer as:

\begin{equation}\label{int}
	E_{\text{int}} = E_{\text{dimer}} - 2\dot E_{\text{monomer}},
\end{equation}

and for the water-peptide system as:

\begin{equation}\label{int}
	E_{\text{int}} = E_{\text{dimer}} - E_{\text{monomer 1}} - E_{\text{monomer 2}}. 
\end{equation}

\subsection{Canonical CC calculations}

CCSD(T) calculations were carried out in Molpro 2022.3.\cite{MOLPRO1, MOLPRO2} %Both counterpoise (CP) corrected corrected results\cite{boys:1970} and the un-counterpoise corrected (raw) results are calculated; the difference between these two values is used as the BSSE error associated with the calculated interaction energies. Half CP (0.5CP) corrected results (the average of the CP and raw results) are also reported. 
%The FP approach means that results reported as CCSD(T)/CBS are, in actuality, often calculated as:
%
%\begin{equation}
%  E_{\text{CCSD(T)}} = E_{\text{SCF}}^{\text{large}}
%    + E_{\text{MP2}}^{\text{large}}
%     + E_{\text{CCSD(T)-MP2}}^{\text{small}} \;.
%\end{equation}

%with 
CBS extrapolations are calculated using the form of Helgaker\cite{halkier:1999} and Halkier,\cite{halkier:1999} as follows:

\begin{equation}\label{helgaker}
	E(XY) = \frac{(X^{3}E_{X} - Y^{3}E_{Y})}{X^{3}- Y^{3}} 
\end{equation}
where \textit{X} and \textit{Y} describe the cardinal number of the basis set and $E_{X}$ and $E_{Y}$ are the total correlation energies in each basis set, respectively. Basis sets employed are the augmented correlation consistent Dunning\cite{dunning:1989} aug-cc-pV\textit{X}Z basis sets with \textit{X}= T, Q, or 5 abbreviated to aV\textit{X}Z. The aVTZ and aVQZ extrapolation is, for example, then denoted as aV\{T, Q\}Z. Basis set extrapolations are only carried out on the correlation energy which are more slowly convergent with BS size while the HF energy is not extrapolated due its faster convergence with BS size.\cite{helgaker:1997, halkier:1999} Where density fitting is employed, JKFIT\cite{weigend:2002} and MPFIT\cite{weigend:2002b} of the same cardinal number as the atomic orbital basis set are used.

For the explicitly correlated calculations, the F12b explicit correlation method\cite{adler:2007, knizia:2009} is used as it systematically converges to the CBS limit.\cite{knizia:2009}
The scaling of the (T) energy is carried out as:
\begin{equation}\label{tstar}
	E_{\mathrm{(T^{*})}} = E_{\mathrm{(T)}} \frac{E_{\text{MP2-F12}}}{E_{\text{MP2}}}
\end{equation}
where all energies are correlation energies, of (T), MP2 and MP2-F12 to, ultimately, give a (T$^{*}$) correlation energy. The scaling factor of the dimer is used for the monomer scaling in both the raw and CP corrected calculations to ensure consistency between calculations and in keeping with previous recommendations.\cite{ma:2018b} It is unclear how these methods should be extrapolated to the CBS limit and, therefore, no basis set extrapolations are carried out for either the canonical or local explicitly correlated methods. 

For the Co.Co.-CCSD(T) calculations, the core region was set to zero and the core-correlated correlation consistent basis sets, aug-cc-pCV\textit{X}Z,\cite{peterson:2002} were used. For the ECP-CCSD(T) calculations, the energy-consistent correlated electron pseudopotentials (eCEPPs) of Trail and Needs were employed.\cite{trail:2017}

\subsection{PNO-LCCSD(T*)-F12b}

Local CC calculations were carried out using (T*) scaled, explicitly correlated pair natural orbital local CCSD(T) (PNO-LCCSD(T*)-F12b).\cite{ma:2018,MOLPRO1, MOLPRO2} Tight domain and pair thresholds\cite{ma:2018b, werner:2023} are used and complete auxiliary basis sets (CABS) corrections\cite{adler:2007, knizia:2009} of the HF energy are utilized. The energy threshold for the local CCSD pair natural orbital domains was set to 0.997, as recommended for intermolecular interactions. % in the Molpro manual. %In the same way as described in eq. \ref{tstar}, the F12 treatment does not extend to the (T) contribution\cite{patkowski:2013} in the PNO-LCCSD(T)-F12b approach and therefore, the triples correlation energy component is scaled by the dimer scaling factor, for all systems within the same basis set size. To indicate the scaling of the (T) correlation energy within this approach, we hereafter refer to this method as PNO-LCCSD(T$^{*}$)-F12b. 

\subsection{SVD-DC-CCSDT+}

The SVD-DC-CCSDT+\cite{kats:2013, kats:2014, rishi:2019, kats:2019, schraivogel:2021,rickert:2025} method is an
approximation to CC theory that uses the distinguishable cluster approximation to remove selected exchange terms from
the CC amplitude equations combined with an SVD treatment of the cluster amplitudes.\cite{hino:2004, kinoshita:2003, lesiuk:2020} SVD-DC-CCSDT+ calculations are carried out in ElemCo.jl.\cite{elemcoil} 

An SVD amplitude threshold of 10$^{-6}$ was used in conjunction with the SVD-(T) correction scheme, as described in Ref. \citenum{lambie:2025} to obtain the SVD-DC-CCSDT+ energies. Importantly, the SVD-DC-CCSDT+ method is used to obtain a post CCSD(T) correction to the energy in the aVTZ basis set, calculated within the frozen core approximation as: %These calculations are carried out using the aVDZ or aVTZ basis set. The FC-CCSD(T) correlation energy in the corresponding basis set size was substracted from the SVD-DC-CCSDT+ correlation energy to obtain a $\delta$SVD-DC-CCSDT+ correction. We then add this to our most accurate CCSD(T) result, the AE-CCSD(T)/aV\{T,Q\}Z basis set to obtain a post CCSD(T) calculation of the correlation energy. 

\begin{equation}
    \Delta \text{SVD-DC-CCSDT+}/\text{aVTZ} = E_{\text{corr.}}(\text{SVD-DC-CCSDT+/aVTZ}) - E_{\text{corr.}}(\text{CCSD(T)/aVTZ}).
\end{equation}

\subsection{DMC calculations}

Wave function optimization and DMC runs have been performed using the
\textsc{casino} code. \cite{needs:2020}
Within DMC we handle the eCEPPs \cite{trail:2017} using the T-move
scheme \cite{casula:2006, casula:2010} as DMC localization
approximation, and we use the size-consistent local-energy limiting
Green's function modifications of Zen \textit{et al}.\@
\cite{zen:2016} to enable accurate, efficient extrapolations to zero
time step.  \cite{lee:2011, vrbik:1986}
Note that, by contrast with the calculations reported in Ref.\
\citenum{shi:2025}, the single-particle orbitals in our SJ wave
function are Gaussian expansions of HF orbitals using the aug-cc-pVTZ
basis set \cite{dunning:1989} instead of B-spline \cite{alfe:2004}
re-expansions of local-density-approximation or Perdew-Burke-Ernzerhof
plane-wave orbitals, which we cusp-correct in AE calculations,
\cite{ma:2005} and we optimize our Jastrow \cite{lopezrios:2012,
drummond:2005} and backflow \cite{lopezrios:2006} parameters using
linear least-squares energy minimization
\cite{toulouse:2007,umrigar:2007} instead of unreweighted variance
minimization.  \cite{schmidt:1990}
Further details are reported in section S2 of the supplementary
information.

\section*{Data availability}

The authors declare that all data supporting the findings of this
study are included in the paper and are available within the paper and
its supplementary information files.
Additional data is available upon reasonable request from the authors.

\bibliography{mpi-fkf}

\section*{Acknowledgments}

Financial support from the Max-Planck Society is gratefully acknowledged. 

\section*{Author contributions statement}

S.L. was responsible for the coupled cluster calculations.
P.L.R. was responsible for the diffusion Monte Carlo calculations.
D.K. oversaw the coupled cluster calculations and provided software
implementation for SVD-DC-CCSDT+ calculations.
A.A. supervised the project.
All authors discussed the results and contributed to the preparation
of the manuscript.
All authors have approved the final manuscript.

\section*{Competing interests}

The authors have no competing interests to declare. 

\end{document}

\maketitle

\section{Additional coupled cluster results}\label{ha_df}

\begin{table}[H]
  \centering 
  \caption{%
    Density-fitted (DF) and heavy-atom only augmented (HA) CCSD(T)
    interaction energies for the AcOH dimer.
    0.5CP-corrected results are reported, with the error being one
    half of the total CP correction.}
  \label{acoh_numbers}
  \begin{tabular}{llc} 
    \hline  
    In text-referral & Treatment	& E\textsubscript{int} [kcal mol$^{-1}$]	\\
    \hline
    DF-CCSD(T) &  HF/DF-aV5Z + MP2/DF-aV\{Q,5\}Z + & \multicolumn{1}{c}{} \\
               & $\Delta$CCSD(T)/aV\{T,Q\}Z        & $-19.41(5)$\\
    HA-CCSD(T) & HF/aVQZ+ CCSD(T)/haV\{T,Q\}Z      & $-19.42(2)$ \\
    \hline
  \end{tabular}
\end{table}

%\section{Alternative basis sets for CCSD(T) calculations}

\textbf{Heavy atom augmented basis sets/HA-CCSD(T)}

A widely-employed basis set in the calculation of non-covalent
interactions is the so-called `heavy atom augmented basis set' whereby
augmentation functions are removed from the H atom basis sets, with
reportedly minimal effect on the interaction energies.
\cite{marshall:2010}
Here, we show results that support this for the AcOH dimer.

\textbf{Density fitted basis sets/FP-CCSD(T)}

%To obtain the FP-CCSD(T) result, we calculate the interaction energy
%of the AcOH dimer using:
%\begin{equation}
%  E_{\text{FP-CCSD(T)}} = E_{\text{SCF}}^{\text{DF-aV5Z}}
%     + E_{\text{MP2}}^{\text{DF-aV\{Q,5\}Z}}
%     + \Delta E_{\text{CCSD(T)-MP2}}^{\text{aV\{T,Q\}Z}}.
%\end{equation}
%where 
Density fitting (DF) is often used to reduce the memory and increase
the speed of quantum chemical calculations. For all density fitted
calculations carried out in this work, the JKFIT\cite{weigend:2002}
and MPFIT\cite{weigend:2002b} basis sets are used are at same cardinal
number as the atomic orbital basis set.
%
%The FP-CCSD(T) and FC-CCSD(T) CP corrected results, calculated using
%slightly different basis sets, obtain interaction energies that
%differ by less than 0.03 kcal mol$^{-1}$, indicating that these two
%approaches converging to the CBS limit. 
%
%The use of HA and DF basis sets do not alter the CC results.

\clearpage

\section{VMC and DMC calculations}

\subsection{Trial wave functions and optimization}

Our basic trial wave function is of the Slater-Jastrow (SJ) form,
\begin{equation}
  \Psi_\text{SJ}({\bf R}) = e^{J({\bf R})} \Psi_\text{S}({\bf R})
  \;,
\end{equation}
with
\begin{equation}
  \Psi_\text{S}({\bf R}) =
    \det[\phi_i({\bf r}_j^\uparrow)  ]
    \det[\phi_i({\bf r}_j^\downarrow)]
    \;,
\end{equation}
where $\bf R$ denotes the set of up- and down-spin electron position
vectors $\{{\bf r}_j^\uparrow\}$ and $\{{\bf r}_j^\downarrow\}$,
$\{\phi_i\}$ are spin-independent single-particle molecular orbitals
and $e^J$ is a Jastrow correlation factor.
In our all-electron (AE) calculations we modify the orbitals close to
the nuclei in order to impose the electron-nucleus Kato cusp
conditions. \cite{ma:2005}

\subsubsection{Jastrow factor}

We use Jastrow factors including isotropic electron-electron,
electron-nucleus, and electron-electron-nucleus terms,
\cite{lopezrios:2012}
\begin{equation}
  \label{eq:jastrow}
  J = \sum_{i<j}^{N_{\rm e}} u_{P_{ij}} (r_{ij})
    + \sum_i^{N_{\rm e}} \sum_I^{N_{\rm n}} \chi_{S_{iI}}(r_{iI})
    + \sum_{i<j}^{N_{\rm e}} \sum_I^{N_{\rm n}}
          f_{T_{ijI}}(r_{ij},r_{iI},r_{jI}) \;,
\end{equation}
which are expressed as natural-power expansions in the relevant
inter-particle distances, \cite{drummond:2005}
\begin{equation}
  \label{eq:jastrow_terms}
  \begin{split}
    u_P(r_{ij})    & = t(r_{ij},L_u)
                     \sum_{k=0}^{n_u} a_k^{(P)} r_{ij}^k \;, \\
    \chi_S(r_{iI}) & = t(r_{iI},L_\chi)
                     \sum_{k=0}^{n_\chi} b_k^{(S)} r_{iI}^k \;, \\
    f_T(r_{ij}, r_{iI}, r_{jI}) & = t(r_{iI},L_f) t(r_{jI},L_f)
                     \sum_{k,l,m=0}^{n_f} c_{klm}^{(T)}
                        r_{ij}^k r_{iI}^l r_{jI}^m \;, \\
  \end{split}
\end{equation}
where $n_u=8$, $n_\chi=8$, and $n_f=3$ are expansion orders, $\{a\}$,
$\{b\}$, and $\{c\}$ are optimizable parameters, $L_u=4.5$,
$L_\chi=4$, and $L_f=4$ are fixed cut-off lengths, $t(r,L) = (1-r/L)^3
\Theta_\text{H}(L-r)$ is a cut-off function, and $\Theta_\text{H}(x)$
is the Heaviside step function.
Indices $P_{ij}$, $S_{iI}$, and $T_{ijI}$ allow the use of different
parameter sets for same- and opposite-spin electron pairs and for
different atomic species on distinct molecules.
Our Jastrow factors for the acetic acid monomer and dimer contain
$196$ free parameters, while those for the water, peptide, and
water-peptide systems contain $136$, $256$, and $376$ parameters,
respectively.

\subsubsection{Backflow transformation}

The Slater-Jastrow-backflow (SJB) wave function is of the form
\begin{equation}
  \Psi_\text{SJB} =
    e^{J({\bf R})}
    \Psi_\text{S} [{\bf X}({\bf R})] \;,
\end{equation}
where the quasiparticle coordinates $\{{\bf x}_i({\bf R})\}$ are of
the inhomogeneous form proposed by L\'opez R{\'\i}os \textit{et al.}\@
\cite{lopezrios:2006} including electron-electron, electron-nucleus,
and electron-electron-nucleus terms,
\begin{equation}
  {\bf x}_i = {\bf r}_i
  + \sum_{j\neq i}^N \eta_{P_{ij}}(r_{ij}) {\bf r}_{ij}
  + \sum_{I}^{N_{\rm n}} \mu_{S_{ij}}(r_{iI}) {\bf r}_{iI}
  + \sum_{j\neq i}^N \sum_{I}^{N_{\rm n}} \left[
      \Phi_{T_{ijI}}(r_{ij}, r_{iI}, r_{jI}) {\bf r}_{ij}
    + \Theta_{T_{ijI}}(r_{ij}, r_{iI}, r_{jI}) {\bf r}_{iI}
    \right] \;,
\end{equation}
which, like the functions in the Jastrow factor, are expressed as
natural-power expansions in the relevant inter-particle distances,
\begin{equation}
  \begin{split}
    \eta_P(r) & = t_P(r,L_\eta) \sum_{k=0}^{n_\eta} c^{(P)}_k r^k \;, \\
    \mu_S(r) & = t_S(r,L_\mu) \sum_{k=0}^{n_\mu} d^{(S)}_k r^k \;, \\
    \Phi_T(r_{ij}, r_{iI}, r_{iJ}) &
    = t_T(r_{iI},L_\Phi) t_T(r_{jI},L_\Phi)
      \sum_{k,l,m=0}^{n_\Phi} \phi^{(T)}_{klm}
      r_{ij}^k r_{iI}^l r_{jI}^m \;, \\
    \Theta_T(r_{ij}, r_{iI}, r_{iJ}) &
    = t_T(r_{iI},L_\Phi) t_T(r_{jI},L_\Phi)
      \sum_{k,l,m=0}^{n_\Phi} \theta^{(T)}_{klm}
      r_{ij}^k r_{iI}^l r_{jI}^m \;, \\
  \end{split}
\end{equation}
where $n_\eta=8$, $n_\mu=8$, and $n_\Phi=2$ are expansion orders,
$\{c\}$, $\{d\}$, $\{\phi\}$, and $\{\theta\}$ are optimizable
parameters, and $L_\eta=4$, $L_\mu=4.5$, and $L_\Phi=4.5$ are fixed
cut-off lengths.
Our backflow functions the acetic acid monomer and dimer contain
$230$ free parameters, while those for the water, peptide, and
water-peptide systems contain $295$, $557$, and $819$ parameters,
respectively.

\subsubsection{Wave function optimization}

We have optimized all of our wave functions using linear least-squares
energy minimization \cite{toulouse:2007,umrigar:2007} using a
correlated-sampling approach with $n_\text{opt}$ VMC-generated
real-space configurations.
For the acetic acid monomer and dimer, and for the water, peptide, and
water-peptide systems we have used $n_\text{opt}=10^6$ for
Slater-Jastrow wave functions.
As detailed in Section \ref{sec:opt_backflow}, our
Slater-Jastrow-backflow wave functions have been optimized with a
number of configurations chosen so as to keep the optimization
uncertainty under a threshold; we have used 
$n_\text{opt}=10^7$ for the acetic acid monomer and dimer,
and $n_\text{opt}=10^7$, $10^6$, and $7.5 \times 10^6$ respectively
for the water, peptide, and water-peptide systems.

\subsection{DMC calculation details}

\subsubsection{DMC time step convergence of SJ-DMC results}

Time-step bias is a smooth function of the DMC time step which is
linear in $\tau$ as $\tau\to 0$, and it is often negligible in energy
differences due to cancellation of errors.
However care must be taken to verify the range of linearity and/or the
degree of cancellation on a case-by-case basis, noting that the choice
of localization scheme significantly affects time-step dependence.

We have chosen to use the T-move localization scheme in our
calculations to avoid the beyond-first-order behavior seen for the
determinant localization approximation (DLA) in the time-step bias
plots of the supplementary information of Ref.\ \citenum{shi:2025}.
It should be noted that for all systems in the S66 set, the
``typical'' time-step range of $\tau=0.01$--$0.04$ a.u.\@ suggested
for ECP systems in Ref.\ \citenum{needs:2020} appears falls outside
the regime where total energies can be regarded to be linear functions
of $\tau$, and therefore this range suggestion should not be followed
blindly when the DLA scheme is used.

\begin{figure}
  \centering
  \includegraphics[width=0.48\textwidth]{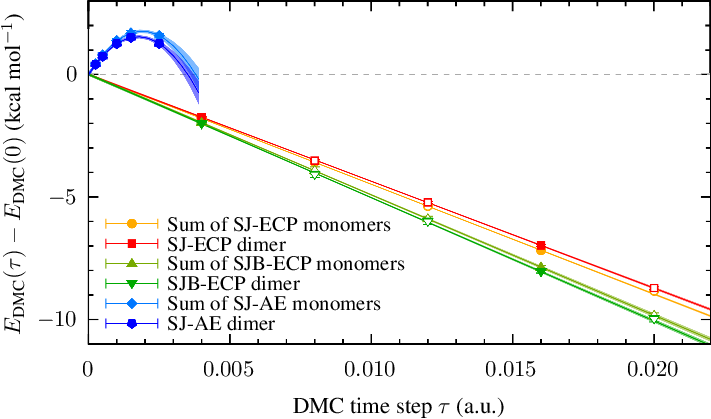}
  \caption{
    Time-step bias on the ECP SJ-DMC and ECP SJB-DMC total energy of
    the AcOH dimer and on twice the total energy of the AcOH monomer,
    along with AE SJ-DMC results for comparison.
    Solid symbols represent data points used in the corresponding fits,
    while open symbols are ignored in fits and shown to verify linear
    trends only.
    Lines represent fits to the data, and shaded areas around them
    represent 95\% confidence intervals on the fit values, while
    errorbars represent 95\% confidence intervals around data points.
    \label{fig:acoh-dimer_dtdmc}
  } 
\end{figure}

\begin{figure}
  \centering
  \includegraphics[width=0.48\textwidth]{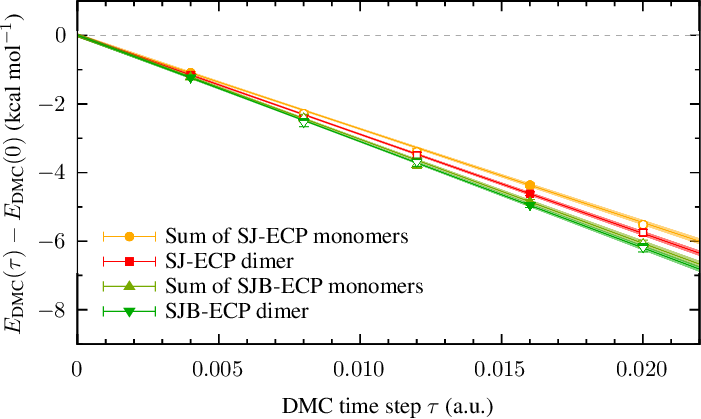}
  \caption{
    Time-step bias on the ECP SJ-DMC and ECP SJB-DMC total energy of
    the water-peptide system and on the sum of the total energies of
    the water and peptide monomers.
    Solid symbols represent data points used in the corresponding fits,
    while open symbols are ignored in fits and shown to verify linear
    trends only.
    Lines represent fits to the data, and shaded areas around them
    represent 95\% confidence intervals on the fit values, while
    errorbars represent 95\% confidence intervals around data points.
    \label{fig:water-peptide_dtdmc}
  } 
\end{figure}

In Fig.\ \ref{fig:acoh-dimer_dtdmc} we plot the time-step dependence
of the total DMC energies obtained for the AcOH monomer and dimer, and
Fig. \ref{fig:water-peptide_dtdmc} shows the time-step dependence of
the total DMC energies obtained for the water, peptide, and
water-peptide systems.
Our ECP SJ-DMC and ECP SJB-DMC results have been obtained by linear
extrapolation to zero time step of the energies obtained at
$\tau=0.004$ and $0.016$ a.u., for efficiency, \cite{lee:2011,
vrbik:1986} and these energies and fits are shown in Figs.\
\ref{fig:acoh-dimer_dtdmc} and \ref{fig:water-peptide_dtdmc}, along
with additional DMC energies at other time steps to provide visual
confirmation that the ECP DMC energy is linear in the time step for
the ECP systems.
Note that the time steps for our ECP calculations were chosen so that
DMC energies using the DLA scheme, should we have chosen to run them,
would have been linearly extrapolatable to zero time step; if we had
targeted the T-move scheme alone from the outset we should have been
able to use significantly larger time steps.

For the AcOH monomer and dimer, Fig.\ \ref{fig:acoh-dimer_dtdmc}
also shows AE SJ-DMC energies, which we extrapolate to zero
time step by fitting all five data points to a second-order
polynomial, and we use this result to test the accuracy of the
ECP approximation.
The two-point linear extrapolation of the ECP SJ-DMC results gives an
interaction energy of $-20.13(6)$ kcal mol$^{-1}$, within uncertainty
of the quadratic extrapolation of the AE SJ-DMC energies of
$-20.29(16)$ kcal mol$^{-1}$, and we conclude that the ECP
approximation does not introduce a significant error.

\subsubsection{DMC population control bias}

We have used a target DMC population of 65536 walkers at the smallest
time step for each system, and have keep the population proportional
to $1/\tau$ at other time steps in order to remove the (likely
negligible) population control bias from our results.
\cite{needs:2020}

\subsection{SJB-DMC optimization uncertainty
            \label{sec:opt_backflow}}

The uncertainty introduced by the stochastic optimization of wave
function parameters is typically ignored in VMC and DMC calculations.
This is partly justified in some cases:\@ the error in the VMC energy
is quadratic in the error in the optimized parameter set when energy
minimization is used, while the SJ-DMC energy of all-electron systems
at zero time step is independent of the parameter values.
However VMC energies of variance-minimized wave functions,
\cite{haupt:2023, filip:2025} VMC expectation values of operators
other than the Hamiltonian, \cite{spink:2016} and beyond-SJ DMC
energies are affected by errors linear in the error in the optimized
parameter set; obtaining accurate values for these requires converging
the results with respect to sample size. \cite{haupt:2023, filip:2025}

\subsubsection{Dependence of the optimization uncertainty on the
VMC-based optimization sample size}

In correlated-sampling optimization, one generates $n_\text{opt}$
random real-space electronic configurations $\{{\bf
R}_i\}_{i=1}^{n_\text{opt}}$ distributed according to an initial wave
function $\Psi({\rm p}; {\bf R})$, where $\bf p$ are the wave function
parameters.
The local energy is evaluated at these configurations,
$E_\text{L}({\bf p},{\bf R}_i) = \Psi^{-1}({\bf p},{\bf R}_i) {\hat
H}({\bf R}_i) \Psi({\bf p},{\bf R}_i)$, and the resulting VMC energy
$E_\text{VMC}({\bf p})$ is computed as the average of the local
energies, which has a standard error proportional to
$n_\text{opt}^{-1/2}$.
The variational energy is then minimized by varying $\bf p$ while
keeping the configurations $\{{\bf R}_i\}_{i=1}^{n_\text{opt}}$ fixed.
The optimizer will clearly only be able to locate the parameters that
minimize the variational energy to the degree that the statistical
uncertainty on the variational energy allows, and thus the the
parameter set ${\bf p}^\prime$ resulting from the optimization
procedure is a random vector which follows a probability distribution
of spread determined by that of the variational energy.
At sufficiently large sample sizes, the central limit theorem applies
to both $E_\text{VMC}({\bf p})$ and ${\bf p}^\prime$, and by error
propagation to leading order, the uncertainty on ${\bf p}^\prime$ is
proportional to that in $E_\text{VMC}({\bf p})$, which in turn makes
it proportional to $n_\text{opt}^{-1/2}$.

The DMC energy is a function of the optimized wave function
parameters, $E_\text{DMC}({\bf p}^\prime)$, and again by error
propagation to leading order, the contribution from the uncertainty on
${\bf p}^\prime$ to the uncertainty on $E_\text{DMC}({\bf p}^\prime)$
is proportional to $n_\text{opt}^{-1/2}$.
Thus, the optimization uncertainty on the SJB-DMC energy is indirectly
dependent on the number of real-space configurations $n_\text{opt}$
used in the VMC-based correlated-sampling optimization process.

\subsubsection{Evaluating the optimization uncertainty}

The most direct, brute-force way of obtaining the optimization
uncertainty, $\sigma_\text{opt}$, for any given system is to perform
multiple independent random instances of wave function optimization
each followed by a DMC run, and evaluating $\sigma_\text{opt}$ as the
standard deviation of the resulting DMC energies.
For the systems in this work we perform $20$ such independent random
instances, and we perform a single DMC run at $\tau=0.016$ a.u.\@
under the assumption that $\sigma_\text{opt}$ is independent of
time step, which we verify below.

In order to ascertain that $\sigma_\text{opt}$ is a well-behaved
function of $n_\text{opt}$, we have evaluated the optimization
uncertainty of the AcOH monomer, $\sigma_\text{opt}^\text{AcOH}$, at
each of four sample sizes $n_\text{opt}$ ranging between $10^4$ and
$10^6$.
Given that the interaction between the AcOH monomers in the dimer is
weak, we take the optimization uncertainty for the AcOH dimer,
$\sigma_\text{opt}^\text{(AcOH)$_2$}$, to be very $\sqrt 2$ times
that of the monomer at the same $n_\text{opt}$.
We have verified this at $n_\text{opt} = 10^5$, where we obtain an
uncertainty ratio of $\sigma_\text{opt}^\text{(AcOH)$_2$} /
\sigma_\text{opt}^\text{AcOH} = 1.72 \pm 0.54$, which well within one
standard deviation from $\sqrt 2$.
We take the total optimization uncertainty on the interaction energy
of the AcOH dimer to be $\sigma_\text{opt} = \sqrt{ (2
\sigma_\text{opt}^\text{AcOH})^2 +
{\sigma_\text{opt}^\text{(AcOH)$_2$}}^2 } = \sqrt 6
\sigma_\text{opt}^\text{AcOH}$, so our target optimization uncertainty
for the interaction energy of $\sigma_\text{opt}=0.071$ kcal
mol$^{-1}$ translates into a target
$\sigma_\text{opt}^\text{AcOH}\approx 0.029$ kcal mol$^{-1}$.

\begin{figure}[H]
  \centering
  \includegraphics[width=0.8\linewidth]{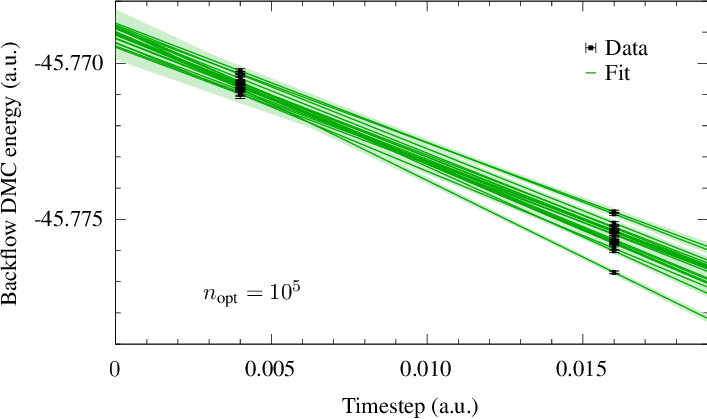}
  \caption{
    ECP SJB-DMC energy as a function of time step for 20 independent
    random instances of wave function optimization at
    sample size $n_\text{opt}=10^5$.
    Lines are linear fits to the data, and shaded areas around them
    represent 95\% confidence intervals on fit values, while
    errorbars represent 95\% confidence intervals around data points.
    \label{fig:acoh-dimer_opt_dtdmc}
  }
\end{figure}

We have also tested the time-step dependence of the optimization
uncertainty by runnning SJB-DMC calculations at $\tau=0.004$ a.u.\ for
each of the $20$ SJB wave functions optimized at $n_\text{opt}=10^5$,
and we plot the corresponding linear extrapolations in Fig.\
\ref{fig:acoh-dimer_opt_dtdmc}.
The standard deviation of the SJB-DMC energies extrapolated to
$\tau=0$ is $0.14$ kcal mol$^{-1}$, which is about half that at
$\tau=0.016$ a.u.\@ of $0.26$ kcal mol$^{-1}$.
We conclude that approximating the optimization uncertainty at
$\tau=0$ by the value obtained at $\tau=0.016$ a.u.\@ provides a safe
overestimation of its true value, and we apply this approximation
throughout.

We compute the optimization sample size by linear extrapolation in
log-log scale using the formula $\log \sigma_\text{opt} = \alpha - 1/2
\log n_\text{opt}$.
For the AcOH molecule we perform a least-squares fit to this formula
using the result from all four sample sizes, and for other systems
where we only perform tests at $n_\text{opt} = 30\,000$ we simply
substitute the values of $\sigma_\text{opt}$ and $n_\text{opt}$ into
the formula and solve for $\alpha$.

For the AcOH monomer and dimer we use $n_\text{opt}=10^7$, which
yields an optimization uncertainty on the interaction energy at
$n_\text{opt}=10^7$ of $\sigma_\text{opt} = 0.060$ kcal mol$^{-1}$,
below our target of $0.071$ kcal mol$^{-1}$.
For the water, peptide, and water-peptide systems we use
$n_\text{opt}=10^7$, $10^6$, and $7.5 \times 10^6$, which yield
optimization uncertainties on the total energies of $0.023$, $0.042$,
and $0.049$ kcal mol$^{-1}$, respectively, and an optimization
uncertainty on the interaction energy of $\sigma_\text{opt} = 0.069$
kcal mol$^{-1}$, again below our target of $0.071$ kcal mol$^{-1}$.

\clearpage

  \section{Geometries}\label{acoh_geom}

\subsection*{AcOH dimer}
    \begin{verbatim}
16

C   -1.061709204   1.297140572   0.292060003
O   -0.358161116   2.270458613   0.531812668
O   -0.589303516   0.094917758   0.003788813
H    0.404435659   0.127722621   0.018411838
C   -2.558427798   1.342549823   0.296257320
H   -2.895997978   2.347464002   0.518316340
H   -2.932889278   1.022390451  -0.672995551
H   -2.937211960   0.644910433   1.039557084
C    2.789348447   1.108419242   0.271183758
O    2.085730082   0.135104754   0.031396156
O    2.316922113   2.310854630   0.558962229
H    1.323133573   2.277956396   0.544561725
C    4.286060895   1.062516496   0.269219363
H    4.623640459   0.061197303   0.031693868
H    4.667559440   1.772869435  -0.460249535
H    4.657577206   1.365211013   1.245274724
\end{verbatim}

\subsection*{Water-peptide}

\begin{verbatim}
	15
	
	O   -0.392018453  -0.384718737   0.076071325
	H   -0.911460851   0.413812040   0.177648774
	H    0.524903820  -0.068484694   0.090511364
	C    2.197705212  -2.245403490  -0.230313251
	H    2.847668055  -3.106515367  -0.363228638
	H    1.516729242  -2.167931431  -1.074178526
	H    1.584688314  -2.384199477   0.656695112
	C    2.952437290  -0.947390613  -0.097719744
	O    2.375721843   0.127904243   0.058868998
	N    4.303070413  -1.044893297  -0.162337713
	H    4.704022041  -1.955427276  -0.291852811
	C    5.171312525   0.107077163  -0.052894631
	H    4.534818395   0.975377606   0.081889976
	H    5.836902032   0.015621965   0.803198250
	H    5.765778248   0.236497652  -0.955153818
\end{verbatim}

\clearpage

\color{black}
\singlespacing
\fontsize{11}{0}
\bibliographystyle{achemso}
\bibliography{mpi-fkf}